\documentclass{ws-procs975x65}

\def\beq{\begin{equation}}
\def\eeq{\end{equation}}

\begin{document}

\title{Wormholes, energy conditions and time machines}

\author{Francisco S. N. Lobo$^{1,*}$ and Diego Rubiera-Garcia$^{2,**}$}

\address{$^{1}$Instituto de Astrof\'{\i}sica e Ci\^{e}ncias do Espa\c{c}o, \\Faculdade de
Ci\^encias da Universidade de Lisboa, \\Edif\'{\i}cio C8, Campo Grande,
P-1749-016 Lisbon, Portugal\\
$^{2}$Departamento de F\'{i}sica Te\'{o}rica and IPARCOS,\\ Universidad Complutense de Madrid, E-28040 Madrid, Spain\\
$^{*}$E-mail: fslobo@fc.ul.pt, $^{**}$E-mail: drubiera@ucm.es}



\begin{abstract}
This report is based on the Parallel Session AT3 ``Wormholes, Energy Conditions and Time Machines'' of the Fifteenth Marcel Grossmann Meeting - MG15, held at the University of Rome ``La Sapienza'' -- Rome, in 2018.
\end{abstract}


\bodymatter


\section{Introduction}

Traditionally, the solutions of General Relativity (GR) are obtained by first considering a plausible distribution of mass and energy, described by a stress-energy tensor, and through the Einstein field equations (EFE) one obtains the spacetime metric. However, one may adopt the reverse philosophy and engineer an interesting metric and then solve the EFE to find the matter sources that sustain the corresponding geometry. In this manner, in wormhole physics, the tunnel-like structure imposes the so-called flaring-out condition at the wormhole throat, the latter being the location of the minimum radius of the geometry. Thus, from this flaring-out condition, 
and using the EFE, in GR one discovers that the null energy condition (NEC) is violated \cite{Morris:1988cz}. In fact, all of the pointwise energy conditions are violated. Another interesting feature of these spacetimes is that they allow ``effective'' superluminal travel, and consequently generate closed timelike curves \cite{Morris:1988tu}, although locally the speed of light is not surpassed. One may mention that, in fact, these solutions are primarily useful as ``gedanken-experiments'' and as a theoretician’s probe of the foundations of GR. We refer the reader to the books \cite{Visser:1995cc,Lobo:2017oab} for an extensive review on the subject.

The AT3 Parallel Session, ``Wormholes, Energy Conditions and Time Machines'', was dedicated to the state-of-art contributions on this fascinating branch of research. In this report, we find it useful to organise the following sections, which reflect the internal schedule that was adopted during the session.

\section{Traversable wormholes: Mathematical structure and applications}



Ever since their revival, three decades ago, in the seminal work of Morris and Thorne \cite{Morris:1988cz,Morris:1988tu}, Lorentzian wormholes in GR have led an uncomfortable existence as they require matter which violates the well known energy conditions. However, in other theories of gravity beyond GR wormholes can indeed exist supported by ``standard matter" satisfying the energy conditions \cite{Harko:2013yb}. 

Sayan Kar illustrated this fact with some known examples and also by explicitly constructing a zero Ricci scalar wormhole in a four dimensional scalar-tensor, on-brane gravity theory arising from the two-brane Randall-Sundrum model with one extra dimension \cite{Aneesh:2018hlp}. If such a wormhole could arise as the end-state of some astrophysical process, its ringdown may be studied using gravitational waves. With this aim, the scalar quasinormal modes in this class of wormholes were obtained and identified as the `breathing mode' associated with gravitational waves in scalar-tensor theories. Moreover, using the frequencies and time constants of the breathing mode as well as the results arising out of the GW150914 event, the size of the wormhole parameters using standard tools were estimated. The analysis suggests that it may be possible to constrain wormholes and distinguish them from black holes using this kind of observations. 



Jutta Kunz considered Ellis-Bronnikov wormholes \cite{homerellis,bronikovWH} immersed in rotating matter in the form of an ordinary complex boson field \cite{Hoffmann:2017vkf,Hoffmann:2018oml}. The resulting wormholes may possess full reflection symmetry with respect to the two asymptotically flat spacetime regions. However, wormhole solutions also arise where the reflection symmetry is broken, and in such a case they always appear in pairs. The properties of these rotating wormholes were analysed and it was shown that their geometry may feature single throats or double throats. The ergoregions and the light ring structure of these wormholes were also discussed.


Kirill Bronnikov analysed spherically symmetric configurations in GR, supported by nonlinear electromagnetic fields with gauge-invariant Lagrangians \cite{Bronnikov:2017sgg} depending on the single invariant $f = F_{\mu\nu} F^{\mu\nu}$. Considering metrics with two independent functions of time, a natural generalization of the class of wormholes previously introduced by Arellano and Lobo with a time-dependent conformal factor was found \cite{Arellano:2006ex}. Such wormholes are shown to be only possible for some particular choices of the Lagrangian function $L(f)$, having no Maxwell weak-field limit. Their time evolution contains cosmological-type singularities. For general non-Maxwell $L(f)$, instead of the usual electric-magnetic duality there exists the so-called FP duality, connecting theories with different $L(f)$; accordingly, for some of the wormhole solutions naturally emerging with magnetic fields, the electric counterparts with the same metric are ill-defined from the viewpoint of the Lagrangian formalism.


Gonzalo Olmo showed that the field equations of metric-affine theories of gravity whose Lagrangian is a nonlinear function of the Ricci tensor can be written in an Einstein form introducing both an auxiliary metric and auxiliary matter fields \cite{Afonso:2018mxn,Afonso:2018hyj}. In this way the resolution of the equations can be done directly from within the framework of GR coupled to nonlinear matter fields. In other words, the nonlinearities of the gravity sector can be transferred to the matter sector. The correspondence with the original nonlinear gravity theory only involves algebraic manipulations. Standard analytical and numerical methods can thus be used to find solutions of these theories starting from GR ones.


Rajibul Shaikh studied shadows cast, under some circumstances, by a certain class of rotating wormholes, and explored the dependence of the shadows on the wormhole spin \cite{Shaikh:2018kfv}. The results were then compared with that of a Kerr black hole. For small spin, the shapes of the shadows cast by a wormhole and a black hole are nearly identical. However, with increasing values of the spin, the shape of a wormhole shadow starts deviating considerably from that of a Kerr black hole. Detection of such a considerable deviation in future observations may possibly indicate the presence of a wormhole. In other words, the results indicate that the wormholes considered in this work, and that have reasonable spin, can be distinguished from a black hole through the observation of their shadows.


João Rosa presented wormhole solutions in the scalar-tensor representation of the generalized hybrid metric-Palatini matter theory, given by a gravitational Lagrangian density $f (R,{\cal R})$, where $R$ is the metric Ricci scalar, and ${\cal R}$ is a Palatini scalar curvature defined in terms of an independent connection \cite{Rosa:2018jwp}. The main interest in the solutions found is that the matter field obeys the NEC everywhere, including the throat and up to infinity, so that there is no need for exotic matter. The wormhole geometry with its flaring out at the throat is supported by the higher-order curvature terms, which can be interpreted as a gravitational fluid. Thus, in this theory, in building a wormhole, it is possible to exchange the exoticity of matter by the exoticity of the gravitational sector.


Sung-Won Kim presented recent work on the exact solution of the cosmological model, in an expanding Friedmann-Lemaitre-Robertson-Walker universe, from the isotropic form of the Morris-Thorne type wormhole \cite{Kim:2018aaw}. Matter, radiation, and the cosmological constant $\Lambda$ were considered as the single component of the universe. For the multi-component universe case,  the $\Lambda$CDM universe with matter and $\Lambda$ was adopted. In the case of the single-component universe, the apparent horizons coincide at a certain time. However, they have two coincidence times in the multi-component universe, which means that they start at early times and finish at late times like the combination of two single-component universes.



Mikhail Volkov applied duality rotations and complexifications to the vacuum Weyl metrics generated by massive rods or by point masses \cite{Gibbons:2017jzk}. As a first step this gives families of prolate and oblate vacuum metrics. Further duality transformations produce a scalar field from the vacuum, which can be either the conventional scalar or the phantom field with negative kinetic energy. This gives rise to large classes of axially symmetric solutions, presumably including all previously known solutions for gravity-coupled massless scalar fields. Particularly interesting are the oblate solutions which, irrespectively of whether they are coupled to a scalar field or not, describe wormholes connecting several asymptotic regions. In the one-wormhole sector they reduce to the ring wormholes in the vacuum case and to the Bronnikov-Ellis wormhole \cite{homerellis,bronikovWH} in the phantom one. The two-wormhole solutions were also studied and it was found that two of their four asymptotic regions are completely regular while two others contain an infinitely long strut along the symmetry axis.


Remo Garattini argued that, since Yukawa corrections to Newtonian potentials appear in some modified theories of gravity, it is worth exploring some aspects of traversable wormholes with a shape function having a Yukawa-like  profile and related generalizations \cite{Garattini:2019wka}. These aspects also include the introduction of some equations of state. More specifically, an equation of state was considered in which the sum of the energy density and radial pressure is proportional to a Yukawa profile. In some cases, it was shown that the specific wormhole solution has an asymptotic behaviour corresponding to a global monopole.


Fayçal Hammad presented the behavior of black hole horizons and wormholes under Weyl conformal transformations \cite{Hammad:2018ldj}. First, a shorter, but more general, derivation of the Weyl transformation of the simple prescription for detecting horizons and wormholes given recently in the literature for spherically symmetric spacetimes was provided. Then, the conformal behavior of black hole horizons and wormholes in more general spacetimes, based on more ``sophisticated'' definitions, was presented. The study showed that black holes and wormholes might always arise in the new frame even if they were absent in the original frame. Moreover, it was shown that the various definitions found in the literature might be transformed into one another under such transformations. Finally, the conformal behavior of the required energy conditions for wormholes was discussed.


Jin Young Kim considered a new approach to construct wormholes without introducing exotic matter in Einstein-Born-Infeld gravity with a cosmological constant \cite{Kim:2016pky}. It was argued that the exoticity of the energy-momentum tensor is not essential to sustain the wormhole. The stability of the new wormholes with a scalar perturbation was also investigated, and it was confirmed that the Breitenlohner-Freedman bound holds in Einstein-Born-Infeld gravity.


Elisa Maggio discussed that gravitational wave astronomy can give us access to the structure of black holes, and to potentially probing microscopic corrections at the horizon scale \cite{Maggio:2018ivz}. A general model of a exotic compact object consisting of a Kerr geometry with a reflective surface near the horizon was investigated. This framework can be applied to thin-shell wormholes when the surface is perfectly reflecting. The stability of these geometries under scalar and electromagnetic perturbations was analysed. It was shown that exotic compact objects with a perfectly reflecting surface are affected by the an ergoregion instability when spinning sufficiently fast. This instability might have a crucial impact on their phenomenology. On the other hand, it was found that a partial absorption at the surface is sufficient to quench the ergoregion instability completely. This finding has important consequences for the viability of exotic compact objects and it suggests that they are not necessarily ruled out by the ergoregion instability.

\section{Energy conditions}



Eleni-Alexandra Kontou presented a theme related to the strong quantum energy inequality and the Hawking singularity theorem \cite{Brown:2018hym}. The latter  concerns matter obeying the strong energy condition (SEC), which means that all observers experience a nonnegative effective energy density  thereby guaranteeing the timelike convergence property. However, for both classical and quantum fields, violations of the SEC can be observed in some of the simplest cases, for example, in the massive Klein-Gordon field. Therefore, there is a need to develop theorems with weaker restrictions, namely, energy conditions averaged over an entire geodesic and quantum inequalities, weighted local averages of energy densities. In fact, lower bounds of the effective energy density were derived in the presence of both classical and quantum scalar fields allowing a nonzero mass and nonminimal coupling to the scalar curvature. In the quantum case these bounds take the form of a set of state-dependent quantum energy inequalities valid for the class of Hadamard states. Finally, it was discussed how the classical and quantum inequalities derived can be used as an assumption to a modified Hawking-type singularity theorem.

\section{Causal structure of spacetime}


Sandipan Sengupta presented, in classical gravity theory, explicit examples of vacuum solutions that admit the possibility of time travel (to the past) through their geodesics \cite{Sengupta:2018ypd}. These geometries are built upon metrics whose determinant can continuously go to zero over some extended region of the spacetime. These solutions to the first order field equations satisfy the energy conditions. One may see the existence of these solutions as a motivation to revisit the status of causality in the formulation of classical gravity.


Vaishak Prasad presented the possibility of resolving the chronology protection problem due to the existence of closed time-like curves in Kerr-Newman black holes using modified gravity \cite{Prasad:2018dcg}. First, the details of the causality violation in the Kerr-Newman spacetime were revisited and quantified. It was then shown that the issue also extends onto two of the modified Kerr-Newman solutions: non-commutative geometry inspired and $f(R)$ gravity solutions. The geodesic connectivity of the causality violating region was discussed in both scenarios and the existence of null geodesics was proved. The possibility of preventing causality violation was also explored. It was shown that although in both models the parameters can be chosen such that the causality violating region is removed, the $f(R)$ case is not consistent with cosmological observations. In the case of the non-commutative geometry inspired solution, it was shown that the chronology protection can be ensured by choosing suitable values for the non-commutative parameter, thereby eliminating the causality violating region and the Cauchy horizon.


Ilia Musco showed that in the gravitational collapse to form black holes, trapping horizons (foliated by marginally trapped surfaces) make their first appearance either within the collapsing matter or where it joins on to a vacuum exterior. Those which then move outwards with respect to the matter have been proposed for use in defining black holes, replacing the global concept of an event horizon, which has some serious drawbacks for practical applications. Results were presented from a study of the properties of both outgoing and ingoing trapping horizons, assuming strict spherical symmetry throughout \cite{Helou:2016xyu}. The causal nature was investigated following two different approaches, namely, one using a geometrical quantity related to expansions of null geodesic congruences, and the other using the horizon velocity measured with respect to the collapsing matter. The models treated are simplified, but do include pressure effects in a meaningful way, and the behavior of the horizon evolution was analysed depending on the initial conditions of energy density and pressure of the collapse.

\section*{Acknowledgments}
FSNL acknowledges support from the Funda\c{c}\~{a}o para a Ci\^{e}ncia e a Tecnologia (FCT) Scientific Employment Stimulus contract with reference CEECIND/04057/2017, and research grants Nos. UID/FIS/04434/2020 and CERN/FIS-PAR/0037/2019.
DRG is funded by the \emph{Atracci\'on de Talento Investigador} programme of the Comunidad de Madrid (Spain) No. 2018-T1/TIC-10431, and acknowledges further support from the Ministerio de Ciencia, Innovaci\'on y Universidades (Spain) project No. PID2019-108485GB-I00/AEI/10.13039/501100011033, the FCT research project No.  PTDC/FIS-PAR/31938/2017, the projects FIS2017-84440-C2-1-P (MINECO/FEDER, EU) and H2020-MSCA-RISE-2017 Grant FunFiCO-777740 and the Edital 006/2018 PRONEX (FAPESQ-PB/CNPQ, Brazil) Grant No. 0015/2019.
The authors also acknowledge the FCT research project No. PTDC/FIS-OUT/29048/2017.

\end{document}